\documentclass[prl,aps,a4paper,nofootinbib,twocolumn]{revtex4}  %

\newif\ifusesec
\usesectrue  
   
\usepackage{graphicx} 
\usepackage{mathrsfs,amsmath,amsfonts,amssymb,multirow}

\newcommand{\beq}{\begin{equation}}
\newcommand{\eeq}{\end{equation}}
\newcommand{\prlsection}[1]{{\it{#1}} ---}

\begin{document}

\title{Analytical determination of the periastron advance in spinning binaries from self-force computations}

\author{Donato \surname{Bini}}
\author{Andrea \surname{Geralico}}
\affiliation{Istituto per le Applicazioni del Calcolo ``M. Picone,'' CNR, I-00185 Rome, Italy}

\date{\today}

\begin{abstract}
We present the first analytical computation of the (conservative) gravitational self-force correction to the periastron advance around a spinning black hole.
Our result is accurate to the second order in the rotational parameter and through the 9.5 post-Newtonian level.
It has been obtained as the circular limit of the correction to the gyroscope precession invariant along slightly eccentric equatorial orbits in the Kerr spacetime. The latter result is also new and we anticipate here the first few terms only of the corresponding post-Newtonian expansion.  
\end{abstract}

\maketitle

Analytical results to a high post-Newtonian (PN) level from gravitational self-force (GSF) theory provide fundamental information for an accurate dynamical description of two-body systems in the extreme-mass-ratio limit, i.e., when one body is much more massive than the other, and in the weak field regime.
Furthermore, such information can be successfully converted into other approximation methods useful for modelling comparable-mass binary inspirals, like the Effective-One-Body (EOB) model \cite{Buonanno:1998gg,Buonanno:2000ef,Damour:2001tu}, which is currently improving thanks also to GSF calculations (see, e.g., Refs. \cite{Damour:2009sm,Tiec:2011ab,Barausse:2011dq,Akcay:2012ea,Tiec:2015cxa}).
In the past few years several gauge-invariant quantities have been computed in the framework of black hole (first-order) perturbation theory, including metric-based quantities, like the redshift factor of the particle's orbit, connection-based quantities, like the precession angle of a gyroscope, and curvature-based quantities, like the tidal invariants, all measured along the world line of the perturbing body, starting from the simplest situation of circular orbits in a Schwarzschild spacetime, later generalized to a Kerr spacetime and to eccentric orbits (see, e.g., Refs. \cite{Detweiler:2008ft,Bini:2013zaa,Kavanagh:2015lva,Shah:2012gu,Bini:2015xua,Kavanagh:2016idg,Bini:2018ylh,Dolan:2013roa,Bini:2014ica,Dolan:2014pja,Bini:2014zxa,Bini:2018dki,Barack:2011ed,Bini:2015bfb,Bini:2016qtx,Bini:2016dvs,Akcay:2016dku,Kavanagh:2017wot,Bini:2018aps,Bini:2018kov}). 

In this Letter we provide the first analytical computation of the (conservative) gravitational self-force correction to the periastron advance around a Kerr black hole, obtained as the circular limit of the correction to the spin-precession invariant along slightly eccentric equatorial orbits (which is original too, but will be presented in a separate work \cite{Bini:2019deltapsi}; we anticipate here the first few terms only of the PN expansion).
Our result is accurate to the second order in the rotation parameter and through the 9.5PN level.
Previous studies on periastron advance in spinning black hole binaries with aligned spins on quasicircular orbits have been performed only through semi-analytical and numerical methods, also taking advantage of available numerical relativity (NR) simulations (see Refs. \cite{Hinderer:2013uwa,Tiec:2013twa,vandeMeent:2016hel} and references therein).
We confirm such numerical predictions, and pave the way for an even more enhanced synergic use of all these approaches, being the periastron advance a key observable to test and improve the modeling of binary dynamics.

\prlsection{Spin-precession invariant}
Let us shortly recall the basic steps underlying the derivation of the spin precession invariant $\psi$ and its first-order GSF correction $\Delta \psi$. 
Consider a binary system consisting of a spinning compact body (of mass $m_1$ and spin $S_1$ such that $|S_1|/m_1^2\ll 1$) and a Kerr black hole (of mass $m_2$ and spin  $S_2$, with $\hat a=S_2/m_2^2$ dimensionless) in the extreme-mass-ratio limit (i.e., $q \equiv \frac{m_1}{m_2} \ll 1$).
According to the GSF approximation scheme, the small body can be considered as following an eccentric geodesic orbit in a suitably regularized perturbed spacetime $g^{\rm R}_{\alpha\beta}=g^{\rm R}_{\alpha\beta}(x^\alpha; m_1,m_2,\hat a)$, through $O(q)$, while its associated spin vector is parallel-transported along it. 
The regularized (R) perturbed metric is decomposed as
\beq
\label{pertmet}
g^{\rm R}_{\alpha\beta}=\bar g_{\alpha\beta}+q \, h^{\rm R}_{\alpha\beta} + O(q^2)\,,
\eeq
where $\bar g_{\alpha\beta}$ is the background (Kerr) spacetime with mass $m_2$ and angular momentum $m_2^2\hat a$, and $q \, h^{\rm R}_{\alpha\beta}$  is the first-order GSF metric perturbation. 
Let us denote by $\Omega_r=2\pi/T_r$  and $\Omega_{\phi}=\Phi/T_r$ the radial and (averaged) azimuthal frequencies, respectively, in the perturbed spacetime, with $T_r$ the radial period and $\Phi$ the accumulated azimuthal angle after a full loop, defining in turn the periastron advance $K=\frac{\Phi}{2\pi}\equiv1+k$.
The spin precession is conveniently measured by the dimensionless quantity 
\beq
\label{psi_def}
\psi=1-\frac{\Psi}{\Phi} 
\,,
\eeq
where $\Psi$ is the precession angle accumulated by the spin vector over one period of radial motion.
The GSF correction $\Delta \psi$ to the spin-precession invariant $\psi=\psi(m_2\Omega_r, m_2\Omega_\phi, \hat a; q)$ can be computed following the procedure outlined in Refs. \cite{Akcay:2016dku,Akcay:2017azq} for fixed values of the orbital frequencies, finally re-expressed in terms of the (inverse) dimensionless semi-latus rectum $u_p=1/p$ and the eccentricity $e$ parametrizing the unperturbed orbit. 

$\Delta \psi$ has been computed in Refs. \cite{Kavanagh:2017wot,Bini:2018aps} in a Schwarzschild spacetime by a small-eccentricity expansion (at $O(e^2)$) and through a high PN order, using the Teukolsky formalism and a PN-expanded metric perturbation in the radiation gauge.
We refer to these works for a detailed account on the calculation of the metric perturbation and related issues, including regularization and metric completion.
We have extended such a calculation to the Kerr case, by computing $\Delta \psi$ to the second order both in the black hole rotational parameter and in the eccentricity of the orbit at the 9.5PN level, namely
\beq
\Delta \psi(u_p, e, \hat a) =\sum_{i,j=0}^2 e^i {\hat a}^j \Delta \psi^{(e^i,a^j)}(u_p)\,.
\eeq
The results will be presented in a forthcoming paper \cite{Bini:2019deltapsi}.
We show below the first few terms only, i.e.,
\begin{widetext}
\begin{eqnarray}
\label{fin_res}
\Delta \psi^{(e^0,a^0)}(u_p)&=&  - u_p+\frac94 u_p^2+\left(\frac{739}{16}+B^{e^0}\right) u_p^3+\left(-\frac{587831}{2880}+\frac{31697}{6144}\pi^2
+C^{e^0}(u_p)
\right) u_p^4 
+O(u_p^5)
\,,\nonumber\\
\Delta \psi^{(e^0,a^1)}(u_p)&=&
-\frac12 u_p^{3/2}-\frac{41}{8} u_p^{5/2}+\left(\frac{237}{32}+B^{e^0}\right) u_p^{7/2}+\left(-\frac{2580077}{5760}+\frac{52225}{6144}\pi^2+C^{e^0}(u_p)
\right) u_p^{9/2}
+O(u_p^{11/2})
\,,\nonumber\\
\Delta \psi^{(e^0,a^2)}(u_p)&=&
-u_p^2+\frac{15}{4} u_p^3+\left(\frac{843}{16}+B^{e^0}\right) u_p^4+\left(-\frac{41161}{2880}+\frac{5155}{1536}\pi^2+C^{e^0}(u_p)
\right) u_p^5
+O(u_p^6)
\,,\nonumber\\
\Delta \psi^{(e^2,a^0)}(u_p)&=& u_p^2+\left(\frac{341}{16}+B^{e^2}\right) u_p^3+\left(-\frac{164123}{480}-\frac{23729}{4096}\pi^2
+C^{e^2}(u_p)
\right)u_p^4 
+O(u_p^5)
\,,\nonumber\\
\Delta \psi^{(e^2,a^1)}(u_p)&=&
-\frac18  u_p^{5/2}+\left(-\frac{59}{16}+B^{e^2}\right) u_p^{7/2}+\left(-\frac{274889}{640}-\frac{39529}{4096}\pi^2+C^{e^2}(u_p)
\right) u_p^{9/2}
+O(u_p^{11/2})
\,,\nonumber\\
\Delta \psi^{(e^2,a^2)}(u_p)&=&
-2 u_p^3+\left(-\frac{13}{4}+B^{e^2}\right) u_p^4+\left(-\frac{65091}{160}-\frac{22037}{2048}\pi^2+C^{e^2}(u_p)
\right) u_p^5
+O(u_p^6)
\,,
\end{eqnarray}
where
\begin{eqnarray}
\label{BCdefs}
B^{e^0}&=& -\frac{123}{64}\pi^2\,,\qquad
C^{e^0}(u_p) = \frac{628}{15}\ln(u_p)+\frac{1256}{15}\gamma+\frac{296}{15}\ln(2)+\frac{729}{5}\ln(3)\,,\nonumber\\
B^{e^2}&=& -\frac{123}{256}\pi^2\,,\qquad
C^{e^2}(u_p) = \frac{268}{5}\ln(u_p)+\frac{536}{5}\gamma+\frac{11720}{3}\ln(2)-\frac{10206}{5}\ln(3)\,.
\end{eqnarray}
\end{widetext}
The spin-independent ($O(a^0)$) terms are already known, but we have included them in Eq. \eqref{fin_res} to show the interesting feature that the same coefficients \eqref{BCdefs} enter the PN expansion of the spin-dependent part, suggesting the following possible (partial) re-summation 
\beq
\frac{u_p^3}{1-\hat au_p^{1/2}} B^{e^0,e^2}\,, \qquad
\frac{u_p^4}{1-\hat au_p^{1/2}} C^{e^0,e^2}(u_p)\,,
\eeq
to this order (only).

\prlsection{Periastron advance}
The GSF correction $\Delta k$ to the periastron advance for quasicircular orbits is related to the difference between the limit for vanishing eccentricity of $\Delta\psi$, i.e., lim$_{e\to0}\Delta\psi$, and the corresponding quantity $\Delta\psi^{\rm circ}$ calculated for circular orbits as follows \cite{Akcay:2016dku,Akcay:2017azq}
\beq
\label{circlimdeltapsi}
\lim_{e\to0}\Delta\psi-\Delta\psi^{\rm circ}=\bar G_\psi \Delta k\,,
\eeq
where the factor $\bar G_\psi$ is given in Eq. (4.34) of Ref. \cite{Akcay:2017azq}.
The circular value $\Delta\psi^{\rm circ}$ has been computed in Ref. \cite{Bini:2018ylh} through the 8 PN order, and improved up to the 9.5 PN order in Ref. \cite{Bini:2019deltapsi}.
Using Eq. \eqref{circlimdeltapsi} the GSF correction $\Delta k$ then turns out to be
\beq
\Delta k(y,\hat a)=\Delta k^{a^0}(y)+\hat a \Delta k^{a^1}(y)+\hat a^2\Delta k^{a^2}(y)\,,
\eeq
in terms of the dimensionless frequency variable $y=(m_2\Omega_\phi)^{2/3}$, where
\begin{eqnarray}
\Delta k^{a^0}(y)&=&
2 y+11 y^2+\left(-\frac{109}{4}+\frac{123}{32}\pi^2\right) y^3+O(y^4)\,,\nonumber\\
\Delta k^{a^1}(y)&=&
C_{1.5}^{a^1,{\rm c}}y^{1.5}+C_{2.5}^{a^1,{\rm c}} y^{2.5}+C_{3.5}^{a^1,{\rm c}} y^{3.5} \nonumber\\
&+&C_{4.5}^{a^1,{\rm c}} y^{4.5}+ (C_{5.5}^{a^1,{\rm c}}+C_{5.5}^{a^1,\ln{}}\ln(y))y^{5.5}\nonumber\\
&+&  (C_{6.5}^{a^1,{\rm c}}+C_{6.5}^{a^1,\ln{}}\ln(y))y^{6.5}+C_{7}^{a^1,{\rm c}}y^7\nonumber\\
&+& (C_{7.5}^{a^1,{\rm c}}+C_{7.5}^{a^1,\ln{}}\ln(y))y^{7.5}+C_{8}^{a^1,{\rm c}}  y^8\nonumber\\
&+& (C_{8.5}^{a^1,{\rm c}}+C_{8.5}^{a^1,\ln{}}\ln(y)+C_{8.5}^{a^1,\ln^2{}}\ln^2(y))y^{8.5}\nonumber\\
&+& C_{9}^{a^1,{\rm c}}   y^9
+(C_{9.5}^{a^1,{\rm c}}+C_{9.5}^{a^1,\ln{}}\ln(y)\nonumber\\
&+& C_{9.5}^{a^1,\ln^2{}}\ln^2(y))y^{9.5}
+O_{\rm ln}(y^{10})
\,,\nonumber\\
\Delta k^{a^2}(y)&=&
C_{2}^{a^2,{\rm c}}y^{2}+C_{3}^{a^2,{\rm c}} y^{3}+C_{4}^{a^2,{\rm c}} y^{4} +C_{5}^{a^2,{\rm c}} y^{5}\nonumber\\
&+& (C_{6}^{a^2,{\rm c}}+C_{6}^{a^2,\ln{}}\ln(y))y^{6}\nonumber\\
&+& (C_{7}^{a^2,{\rm c}}+C_{7}^{a^2,\ln{}}\ln(y))y^{7} +C_{7.5}^{a^2,{\rm c}} y^{7.5}\nonumber\\
&+& (C_{8}^{a^2,{\rm c}}+C_{8}^{a^2,\ln{}}\ln(y))y^{8}+C_{8.5}^{a^2,{\rm c}}  y^{8.5}\nonumber\\
&+& (C_{9}^{a^2,{\rm c}}+C_{9}^{a^2,\ln{}}\ln(y)+C_{9}^{a^2,\ln^2{}}\ln^2(y))y^{9}\nonumber\\
&+& C_{9.5}^{a^2,{\rm c}}  y^{9.5}
+O_{\rm ln}(y^{10})
\,,
\end{eqnarray}
with coefficients

\begin{widetext}

\begin{eqnarray} 
C_{1.5}^{a^1,{\rm c}}&=&1
\,,\qquad
C_{2.5}^{a^1,{\rm c}}=\frac{11}{6} 
\,,\qquad
C_{3.5}^{a^1,{\rm c}}=\frac{385}{24}
\,,\qquad
C_{4.5}^{a^1,{\rm c}}=\frac{134813}{144}-\frac{3655}{96}\pi^2
\,,\nonumber\\
C_{5.5}^{a^1,{\rm c}}&=&\frac{1099411769}{86400}-\frac{1345253}{1152}\pi^2+\frac{112528}{45}\gamma
-\frac{8816}{45}\ln(2)+\frac{25272}{5}\ln(3)
\,,\qquad
C_{5.5}^{a^1,\ln{}}=\frac{56264}{45}
\,,\nonumber\\
C_{6.5}^{a^1,{\rm c}}&=&\frac{19713084823}{57600}-\frac{13220837}{384}\pi^2+\frac{1384072}{105}\gamma+\frac{1237496}{105}\ln(2)
+\frac{128304}{5}\ln(3)
\,,\qquad
C_{6.5}^{a^1,\ln{}}=\frac{701108}{105}
\,,\nonumber\\
C_{7}^{a^1,{\rm c}}&=&\frac{24421972}{4725}\pi
\,,\nonumber\\
C_{7.5}^{a^1,{\rm c}}&=&\frac{428677301726249}{130636800}-\frac{34677235519}{98304}\pi^2+\frac{130243318}{2835}\gamma-\frac{567286802}{2835}\ln(2)+\frac{2240703}{10}\ln(3)
\nonumber\\
&&
+\frac{48828125}{1134}\ln(5)+\frac{513781541}{131072}\pi^4
\,,\qquad
C_{7.5}^{a^1,\ln{}}= \frac{67738931}{2835}
\,,\nonumber\\ 
C_{8}^{a^1,{\rm c}} &=& \frac{720335276}{33075}\pi
\,,\nonumber\\
C_{8.5}^{a^1,{\rm c}}&=&\frac{2478301696045232739841}{116181838080000}-\frac{3652164831352613}{1486356480}\pi^2+\frac{15963954477931}{16372125}\gamma-\frac{14852430750587}{16372125}\ln(2)\nonumber\\
&&
+\frac{3154801921389}{1078000}\ln(3)-\frac{689453125}{11088}\ln(5)+\frac{58977748388749}{1509949440}\pi^4+\frac{5017408}{45}\zeta(3)-\frac{311603968}{4725}\gamma^2\nonumber\\
&&
-\frac{402564608}{4725}\gamma\ln(2)-\frac{31131216}{175}\gamma\ln(3)+\frac{19361024}{4725}\ln(2)^2-\frac{31131216}{175}\ln(3)\ln(2)-\frac{15565608}{175}\ln(3)^2
\,,\nonumber\\
C_{8.5}^{a^1,\ln{}}&=&\frac{16271988057631}{32744250}-\frac{311603968}{4725}\gamma-\frac{201282304}{4725}\ln(2) -\frac{15565608}{175} \ln(3)
\,,\qquad
C_{8.5}^{a^1,\ln^2{}}=-\frac{77900992}{4725}
\,,\nonumber\\
C_{9}^{a^1,{\rm c}}&=&\frac{343426024814}{3274425}\pi
\,,\nonumber\\
C_{9.5}^{a^1,{\rm c}}&=&-\frac{11476098516297075424765007}{9773732583014400}-\frac{191705333547482269}{14863564800}\pi^2+\frac{88342652436678389}{11918907000}\gamma\nonumber\\
&&
-\frac{40688575776828179}{11918907000}\ln(2)+\frac{516579668986977}{28028000}\ln(3)+\frac{1983860546875}{2018016}\ln(5)+\frac{11529792238433}{34749000}\ln(7)\nonumber\\
&&
+\frac{107766811323369547}{8053063680}\pi^4 +\frac{92591392}{105}\zeta(3)-\frac{3563997904}{11025}\gamma^2-\frac{41292822752}{33075}\gamma\ln(2)-\frac{102281616}{175}\gamma\ln(3)\nonumber\\
&&
-\frac{59198785616}{33075}\ln(2)^2-\frac{102281616}{175}\ln(3)\ln(2)-\frac{51140808}{175}\ln(3)^2
\,,\nonumber\\
C_{9.5}^{a^1,\ln{}}&=&\frac{90459236975659301}{23837814000}-\frac{3563997904}{11025}\gamma -\frac{4154391152}{6615}\ln(2)
 -\frac{51140808}{175} \ln(3) 
\,,\nonumber\\
C_{9.5}^{a^1,\ln^2{}}&=&-\frac{182384708}{2205}
\,,
\end{eqnarray}
and
\begin{eqnarray}
C_{2}^{a^2,{\rm c}}&=&-1
\,,\qquad
C_{3}^{a^2,{\rm c}}=-\frac{55}{2} 
\,,\qquad
C_{4}^{a^2,{\rm c}}= -\frac{19399}{72} 
\,,\qquad
C_{5}^{a^2,{\rm c}}=-\frac{446447}{144}+\frac{27919}{1024}\pi^2\,,
\nonumber\\
C_{6}^{a^2,{\rm c}}&=&-\frac{186587519}{5760}+\frac{2013857}{3072}\pi^2-\frac{4112}{3}\gamma-\frac{3088}{15}\ln(2)-\frac{12393}{5}\ln(3)
\,,\qquad
C_{6}^{a^2,\ln{}}=-\frac{2056}{3}
\,,\nonumber\\
C_{7}^{a^2,{\rm c}}&=&\frac{53325660193}{3628800}-\frac{37838321197}{1769472}\pi^2-\frac{5670004}{189}\gamma
+\frac{1186988}{189}\ln(2)-\frac{2418903}{35}\ln(3)
\,,\qquad
C_{7}^{a^2,\ln{}}=-\frac{2835002}{189}
\,,\nonumber\\
C_{7.5}^{a^2,{\rm c}}&=&-\frac{759272}{315}\pi
\,,\nonumber\\
C_{8}^{a^2,{\rm c}}&=&\frac{652997684751773}{101606400}-\frac{247887099097913}{309657600}\pi^2-\frac{168677140}{567}\gamma
+\frac{299044108}{2835}\ln(2)-\frac{423860931}{560}\ln(3)\nonumber\\
&&
-\frac{9765625}{432}\ln(5)+\frac{3505784063}{33554432}\pi^4-\frac{17376}{5}\zeta(3)
\,,\qquad
C_{8}^{a^2,\ln{}}=-\frac{434074618}{2835}
\,,\nonumber\\
C_{8.5}^{a^2,{\rm c}}&=&-\frac{494597581}{7938}\pi
\,,\nonumber
\end{eqnarray}
\begin{eqnarray}
C_{9}^{a^2,{\rm c}}&=&\frac{8584168680144409717}{140826470400}-\frac{224107660142701427}{26424115200}\pi^2-\frac{5563865373761}{2182950}\gamma
+\frac{35170547468123}{10914750}\ln(2)  \nonumber\\
&&
-\frac{273231294255}{34496}\ln(3)-\frac{191658203125}{399168}\ln(5)+\frac{231263099856899}{2013265920}\pi^4-\frac{1367824}{15}\zeta(3)+\frac{9688208}{315}\gamma^2
\nonumber\\
&&
+\frac{66970016}{1575}\gamma\ln(2)+\frac{2808108}{35}\gamma\ln(3)+\frac{3572944}{1575}\ln(2)^2+\frac{2808108}{35}\ln(3)\ln(2)+\frac{1404054}{35}\ln(3)^2
\,,\nonumber\\
C_{9}^{a^2,\ln{}}&=&-\frac{5758445969201}{4365900}+\frac{9688208}{315}\gamma+\frac{33485008}{1575}\ln(2)+\frac{1404054}{35}\ln(3)
\,,\qquad
C_{9}^{a^2,\ln^2{}}=\frac{2422052}{315}
\,,\nonumber\\
C_{9.5}^{a^2,{\rm c}}&=&-\frac{14973550314493}{26195400}\pi
\,.
\end{eqnarray}

\end{widetext}
The Schwarzschild term $\Delta k^{a^0}$ is fully known up to the same PN level in terms of the EOB function $\rho$ \cite{Damour:2009sm,Bini:2016qtx}.
The agreement of the spin-dependent part with the PN expectation is shown in Ref. \cite{Bini:2019deltapsi}.

In order to compare our analytical result with existing numerical studies \cite{Tiec:2013twa,vandeMeent:2016hel} we construct the related function $W=K^{-2}=\bar W+qW_{\rm GSF}+O(q^2)$, with $\bar W$ the corresponding Kerr background value, expressed in terms of the variable $x=(M\Omega_\phi)^{2/3}=y(1+q)^{2/3}$, with $M=m_1+m_2$. 
The linear in mass ratio correction $W_{\rm GSF}$ was estimated in Ref. \cite{Tiec:2013twa} by combining the 3.5 PN prediction and available NR simulations for single-spin binary black hole systems with different mass ratios between $1/1.5$ and $1/8$ and spin $\hat a=-0.5$. The GSF correction was then extracted by a fitting procedure in the frequency range $0.012<M\Omega_\phi< 0.036$ (see Eqs. (37)--(41) there).
The correction to the periastron advance was instead calculated in Ref. \cite{vandeMeent:2016hel} within the same GSF framework adopted here, but the solution to the first order perturbation equations and the metric reconstruction procedure were carried out through numerical methods. 
The function $W_{\rm GSF}$ is given by Eq. (4) there, in terms of the numerically computed (averaged) components of the gravitational self-force along the orbit.

\begin{figure}
\includegraphics[scale=0.4]{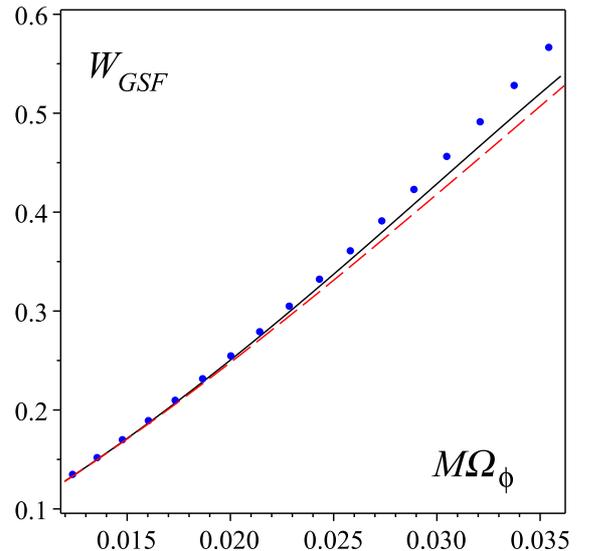}
\caption{\label{fig:1}
The GSF correction $W_{\rm GSF}$ to $W$ as a function of the dimensionless frequency $M\Omega_\phi$ is shown for the value $\hat a=-0.5$ of the black hole spin parameter in comparison with the numerical values (dots) obtained in Ref. \cite{vandeMeent:2016hel} (see Eq. (4) there) as well as the fitting function (dashed curve) of Ref. \cite{Tiec:2013twa} (see Eq. (41) there).  
}
\end{figure}

We show in Fig. \ref{fig:1} the behavior of $W_{\rm GSF}$ in terms of the dimensionless azimuthal frequency $M\Omega_\phi$ for the value $\hat a=-0.5$ of the black hole rotation parameter.
At low frequencies, where the gravitational field is weak enough, all methods agree.
As the frequency increases, i.e., as the strong field regime is approached, our analytical curve maintains closer to the numerical one of Ref. \cite{vandeMeent:2016hel} than the NR-based estimate of Ref. \cite{Tiec:2013twa}, even if the value of the black hole rotational parameter is not so small, being our result accurate up to the second order in spin only.
We also provide in Table \ref{table:1} a comparison between the numerical values of $W_{\rm GSF}$ of Ref. \cite{vandeMeent:2016hel} and our analytical result for a smaller value of $\hat a=0.1$, as an example, and the whole available range of frequencies, showing a better agreement.


\begin{table}[t]
  \caption{
\label{table:1} 
Comparison between the numerical values of $W_{\rm GSF}$ of Ref. \cite{vandeMeent:2016hel} and our analytical result for $\hat a=0.1$ and different values of the frequency.}
  \begin{center}
    \begin{ruledtabular}
      \begin{tabular}{|c|c|}
\hline 
$M\Omega_\phi$ & $W_{\rm GSF\,,num}-W_{\rm GSF\,,analytic}$ \cr
\hline 
0.9999$\times10^{-3}$&  -4.4453$\times10^{-9}$ \cr  
0.2828$\times10^{-2}$&  -5.8628$\times10^{-8}$ \cr  
0.5193$\times10^{-2}$&  -2.7953$\times10^{-7}$ \cr  
0.7994$\times10^{-2}$&  -8.1643$\times10^{-7}$ \cr  
0.1117$\times10^{-1}$&  -0.1384$\times10^{-5}$ \cr  
0.1468$\times10^{-1}$&  8.3614$\times10^{-7}$ \cr  
0.1849$\times10^{-1}$&  0.1756$\times10^{-4}$ \cr  
0.2258$\times10^{-1}$&  0.8468$\times10^{-4}$ \cr  
0.2693$\times10^{-1}$&  0.2931$\times10^{-3}$ \cr  
0.3152$\times10^{-1}$&  0.8442$\times10^{-3}$ \cr  
0.3635$\times10^{-1}$&  0.2139$\times10^{-2}$ \cr  
0.4140$\times10^{-1}$&  0.4910$\times10^{-2}$ \cr  
0.4665$\times10^{-1}$&  0.1040$\times10^{-1}$ \cr  
0.5211$\times10^{-1}$&  0.2060$\times10^{-1}$ \cr  
0.5776$\times10^{-1}$&  0.3847$\times10^{-1}$ \cr  
0.6359$\times10^{-1}$&  0.6829$\times10^{-1}$ \cr  
0.6960$\times10^{-1}$&  0.1160 \cr
\end{tabular}
\end{ruledtabular}
\end{center}
\end{table}

\prlsection{Discussion}
We have presented here the first analytical computation of the conservative part of the gravitational self-force correction to the periastron advance around a Kerr black hole, accurate to the second order in the rotational parameter and to the 9.5PN level.
Our theoretical prediction satisfactorily agrees with existing numerical data in the weak field regime.
We have also anticipated the first PN terms of the GSF correction to the precession angle of a gyroscope moving along a slightly eccentric equatorial orbit in a Kerr spacetime. 
We expect fruitful synergies informing other formalisms, e.g., the EOB model, as well as further improvements in the modeling of the dynamics of eccentric spinning binaries.

\end{document}